# Process Voltage Temperature Variability Estimation of Tunneling Current for Band-to-Band-Tunneling based Neuron

Shubham Patil, Anand Sharma, Gaurav R, Abhishek Kadam, Ajay Kumar Singh, Sandip Lashkare, Nihar Ranjan Mohapatra, and Udayan Ganguly, Senior Member, IEEE

*Abstract* — Compact and energy-efficient Synapse and Neurons are essential to realize the full potential of neuromorphic computing. In addition, a low variability is indeed needed for neurons in Deep neural networks for higher accuracy. Further, process (P), voltage (V), and temperature (T) variation (PVT) are essential considerations for low-power circuits as performance impact and compensation complexities are added costs. Recently, band-to-band tunneling (BTBT) neuron has been demonstrated to operate successfully in a network to enable a Liquid State Machine. A comparison of the PVT with competing modes of operation (e.g., BTBT vs. sub-threshold and above threshold) of the same transistor is a critical factor in assessing performance. In this work, we demonstrate the PVT variation impact in the BTBT regime and benchmark the operation against the subthreshold slope (SS) and ON-regime ($I_{ON}$) of partially depleted-Silicon on Insulator MOSFET. It is shown that the On-state regime offers the lowest variability but dissipates higher power. Hence, not usable for low-power sources. Among the BTBT and SS regimes, which can enable the low-power neuron, the BTBT regime has shown ~3× variability reduction ($\sigma_{I_D}/\mu_{I_D}$) than the SS regime, considering the cumulative PVT variability. The improvement is due to the well-known weaker P, V, and T dependence of BTBT vs. SS. We show that the BTBT variation is uncorrelated with mutually correlated SS & $I_{ON}$ operation – indicating its different origin from the mechanism and location perspectives. Hence, the BTBT regime is promising for low-current, low-power, and low device-to-device variability neuron operation.

## I. INTRODUCTION

The evolution of the Internet of Things (IoT) led to an explosion in data management complexity. It raised the requirement for data-centric processing to increase efficiency [1]. Among the different proposed architectures to improve the efficiency [2]–[6], neuromorphic computing inspired from the human brain is gaining significant interest. computing. Spiking Neural Networks (SNNs), a 3rd generation neural

This work was supported in part by the Department of Science and Technology (DST), India, the Indian Institute of Technology Bombay Nano-Fabrication (IITBNF) facility IIT Bombay, and the Centre for Excellence (CEN), IIT Bombay. The authors would like to thank Global Foundries for providing wafers.

Shubham Patil, Anand Sharma, Gaurav R, Abhishek Kadam, Ajay Kumar Singh, Sandip Lashkare, and Udayan Ganguly are with the Department of Electrical Engineering, Indian Institute of Technology Bombay, India (e-mail: shubhampatil2.107@gmail.com, udayan@ee.iitb.ac.in).
Nihar R. Mohapatra is with the Department of Electrical Engineering, IIT Gandhinagar, Gujarat 382355, India (e-mail: nihar@iitgn.ac.in).

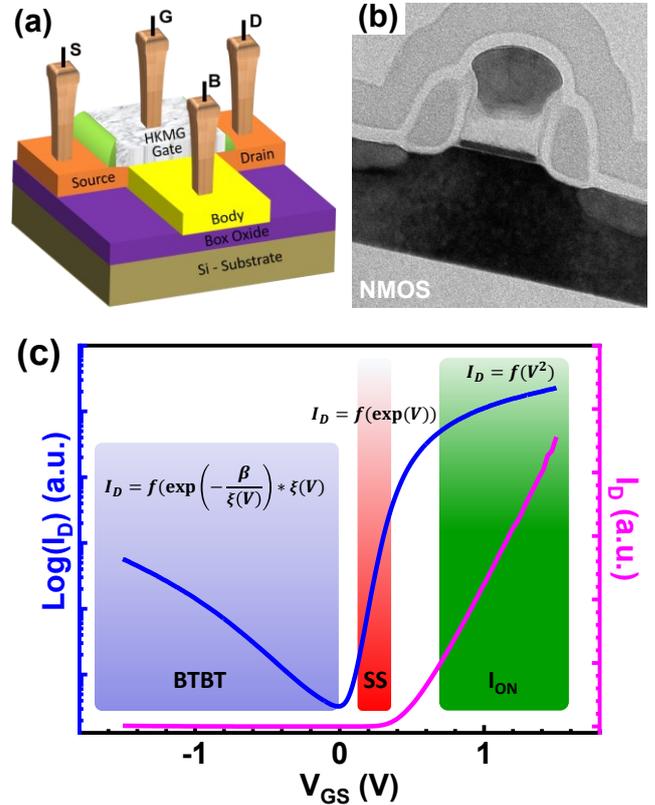

Figure 1 **SOI MOSFET.** (a) Schematic representation and (b) TEM image of the 32 nm PD-SOI MOSFET fabricated at Global Foundries [34]–[36]. **(c)** Experimentally measured transfer characteristics ($I_D$ – $V_{GS}$) of the SOI MOSFET. The three operating regimes of MOSFET (i) BTBT, (ii) SS, and (iii) $I_{ON}$ are highlighted in the figure, which is generally used for neuron demonstration. The current equations are referred from [24]–[27].

network, offer energy and area-efficient realization of the human brain [7]. In addition to the synapse, compact and energy-efficient neurons are required to realize large SNN networks.

The neuromorphic community explored various conventional Silicon-based and non-Silicon-based neurons [8]–[10]. However, the existing Si-based neurons [11]–[16] suffer from the large area required for implementation. Further, the typical ON-regime ($I_{ON}$) based CMOS neurons require a dedicated on-chip capacitor. To reduce the drive current and the capacitance, SS-based operation was used [23]. However, the current in the SS regime can't be reduced below a certain threshold because of high variability. To reduce the variability in the SS regime, the designer has to trade-off with area [17]–[19]. To reduce the on-chip capacitor, the body effect in the SOI transistor was used with the current generated by Impact Ionization (II) [20]–[22]. The experimentally demonstrated neuron power consumption is high due to its operation in the ON-regime with high $I_{ON}$. To reduce power and improve energy efficiency, II was replaced with band-to-band tunneling (BTBT) to reduce current [7]. Thus, a compact low-power integrator for a neuron was proposed. This technology was based on standard SOI technology, so further integration was supported naturally. First, the BTBT integrator was implemented in a full neuron by adding a CMOS-based threshold and reset circuit. Second, a bank of 36 neurons was integrated with synapses to demonstrate a liquid state machine (LSM) that enables spoken word detection [23].

In addition to compactness and low power, neuron variability is indeed one of the major concerns. The integration time and the threshold at which the neuron starts firing are greatly affected by variability. The variation in integration time leads to variation in the spiking frequency of neurons. As we continue to reduce the power of the neuron, it becomes more vulnerable to voltage and temperature variability, which can cause errors in the performance of any real-time application, such as classification and optimization tasks. Hence, Process, Voltage, and Temperature (PVT) variability are essential considerations for low-power neural networks, as performance impact and compensation complexities due to variability are added costs. For (a) the commercialization of the chip for applications or product-level demonstration and (b) improving the circuit models, the experimental process, voltage, and temperature (PVT) variability estimation is critical for the statistically less-explored BTBT regime of operation.

In this paper, we characterize and compare PVT variability in BTBT current in the same SOI transistor with the traditional current sources, e.g., Subthreshold Slope (SS) regime and ON-regime ($I_{ON}$) (Fig. 1). The effect of process variation is measured by statistically testing devices of the same geometry. The effect of voltage variation is measured by the voltage sensitivity of the currents. The temperature effects are measured by varying the temperature of measurements. We compare the variability in BTBT, SS, and $I_{ON}$ regimes as a function of the operating current to show that BTBT current source is promising for low current applications in the spiking neuron.

## II. PVT VARIABILITY DISCUSSION

We have performed a first-order comparative analysis of these operating regimes for PVT variability using the mathematical models available in the literature [24]–[27] as listed below (1)-(8). The drain current equation using the source-referenced model in the strong inversion regime (above threshold or $I_{ON}$ regime) and weak inversion (Sub-threshold Slope regime) is given by (1) and (2), respectively [24]. The current in the BTBT regime is given by (3) [26]. The impact of PVT variability on current in the three operating regimes is discussed below:

*Process Variability:* For simplifying the analysis, the process variation impact is only considered in the threshold voltage ($V_T$), gate oxide thickness ($t_{ox}$) and device dimensions (width (W)/gate length (L)). It can be observed that the $I_{ON}$ regime current (1) shows quadratic dependence on $V_T$, whereas the SS regime current (2) shows exponential dependence. Further, both regimes show direct dependence on the device dimension (W/L) and $t_{ox}$ variation. On the other hand, the BTBT regime current (3) is independent of $V_T$. It depends on variation in the drain-gate overlap, $t_{ox}$, defects and random dopant fluctuation at drain/channel interface, doping abruptness at drain junction, gate structure at the drain end (e.g., corner rounding, etc.), and the source/drain resistance ($R_{SD}$), causing a change in the tunneling distance/barrier at the junction [24], [27].

*Voltage Variability:* The $I_{ON}$ regime shows quadratic dependence on voltage, whereas the SS regime shows exponential dependence. The BTBT regime current depends primarily exponentially on the inverse of the electric field across the tunneling distance at the channel-drain junction. The electric field depends on the voltage across the junction (4) [26], [27].

*Temperature Variability:* With the increase in the temperature, (i) the $I_{ON}$ regime current decreases due to an increase in phonon scattering [24] and shows sub-quadratic dependence, (ii) the SS regime shows an exponential increase in current due to a decrease in $V_T$ in accordance with Fermi-Dirac (FD) statistics and (iii) the BTBT regime current depends on the change in the tunneling distance with the decrease in the bandgap which is a weak dependence [26].

$$I_{D,I_{ON}} = \frac{W}{L}\mu_{eff}(T)C_{ox}\frac{(V_{GS}-V_T(T))^2}{2n(T)} \quad (1)$$

$$I_{D,SS} = \frac{W}{L}\mu_{eff}C_{ox}(n(T)-1)\left(\frac{KT}{q}\right)^2$$
$$* e^{\left(\frac{q(V_{GS}-V_T(T))}{n(T)KT}\right)}\left(1-e^{\frac{qV_{DS}}{KT}}\right) \quad (2)$$

$$I_{D,BTBT} = \frac{\sqrt{2m^*}q^3\xi(T)V_R}{4\pi^3\hbar^2 E_g^{\frac{1}{2}}(T)}e^{\left(\frac{-4\sqrt{2m^*}E_g^{\frac{3}{2}}(T)}{3q\xi(T)\hbar}\right)} \quad (3)$$

$$\xi(T) = \sqrt{\frac{2qN_AN_D(V_R+V_{bi}(T))}{\epsilon_{si}}} \quad (4)$$

$$V_T(T) = V_{FB}(T) + \phi_B(T) + \gamma\sqrt{2*\phi_B(T)} \quad (5)$$

$$V_{FB}(T) = \phi_M - \left(\chi_s + \frac{E_g(T)}{2q} + \phi_B(T)\right) \quad (6)$$

$$\phi_B(T) = \frac{KT}{q}\ln\left(\frac{N_A}{n_i(T)}\right), n(T) = 1 + \frac{\gamma}{2\sqrt{\phi_B(T)}} \quad (7)$$

$$\gamma = \frac{\sqrt{2\epsilon_{Si}qN_A}}{C_{ox}}, \beta = -4\sqrt{2m^*}E_g^{\frac{3}{2}}(T) \quad (8)$$

where, $I_{D,I_{ON}}, I_{D,SS}, I_{D,BTBT}$ represent drain current in above threshold ($I_{ON}$ regime), sub-threshold slope (SS) and band-to-band-tunnelling (BTBT) regime, respectively, $W$ is the channel width, $L$ is the gate length, $\mu_{eff}$ is the effective channel mobility, $C_{ox}$ is the gate capacitance per unit area, $V_{GS}$ is the gate-to-source voltage, $V_{DS}$ is the drain-to-source voltage, $V_T$ is the threshold voltage, $\phi_M$ is the metal workfunction, $\chi_s$ is the electron affinity, $\phi_B$ is the bulk potential, $\gamma$ is the body-bias coefficient, $K$ is the Boltzmann constant, $T$ is the temperature, $N_A$ is the channel doping concentration, $N_D$ is the source/drain doping concentration, $m^*$ is the effective mass, $q$ is the electronic charge, $\xi$ is the electric field, $V_R$ is the reverse bias voltage across the junction, $V_{bi}$ is the built-in potential, $\epsilon_{si}$ is the relative permittivity of Silicon, $E_g$ is the bandgap, $\hbar$ is the Planck's constant and $n_i$ is the free carrier concentration.

TABLE I. BAND DIAGRAM AND FIRST-ORDER COMPARATIVE ANALYSIS OF THE BTBT, SS, AND $I_{ON}$ REGIME FOR THE PROCESS, VOLTAGE, AND TEMPERATURE VARIABILITY [24]–[27].

| Comparison | BTBT | SS | $I_{ON}$ |
|---|---|---|---|
| Band diagrams | $I_{D,BTBT}$ band diagram with $E_G$ | $I_{D,SS}$ band diagram | $I_{D,I_{ON}}$ band diagram |
| Process (P) | $I_D$: $V_T$ independent, inherent variability dependent | $I_D \propto \frac{W}{L}$<br>$I_D \propto e^{\frac{q(V_{GS}-V_T)}{KT}}$<br>Strong exponential dependence on $V_T$ | $I_D \propto \frac{W}{L}$<br>$I_D \propto (V_{GS}-V_T)^2$<br>Quadratic dependence on $V_T$ |
| Voltage (V) | $I_D \propto \xi(V) \, e^{-\frac{\beta}{\xi(V)}}$<br>Weak exponential dependence | $I_D \propto e^{\frac{q(V_{GS}-V_T)}{KT}}$<br>Exponential dependence | $I_D \propto (V_{GS}-V_T)^2$<br>Quadratic dependence |
| Temperature (T) | $I_D \propto \frac{\xi(T)}{E_g^{\frac{1}{2}}(T)} e^{\left(-\frac{E_g^{\frac{3}{2}}(T)}{\xi(T)}\right)}$<br>Weak exponential dependence | $I_D \propto e^{\frac{q(V_{GS}-V_T(T))}{KT}}$<br>Exponential dependence | $I_D \propto \mu_{ph}(T^{-\frac{3}{2}})$<br>Sub-quadratic dependence |

The reduced form of the drain current equations with dependence on the PVT variability are summarized in Table I. To summarize the first-order analysis, the $I_{ON}$ and SS regime shows direct dependence for the PVT variability through the device dimension, threshold voltage, and mobility degradation [24], [26], [28]–[31]. It can be observed that the SS regime shows much stronger dependence than the $I_{ON}$ regime. However, the BTBT regime uses different device physics [24], [25], [32], [33], which leads to potentially weaker PVT variability compared to the SS regime. However, this potential benefit needs to be estimated experimentally. Hence, we have performed an experimental variability study on the SOI MOSFET in these three operating regimes with performance benchmarking for the PVT variability.

### III. DEVICE AND MEASUREMENT DETAIL

The 32 nm PD-SOI MOSFET devices used in this work are fabricated at GlobalFoundries [34]–[36]. The schematic cross-section and the TEM image of the PD-SOI MOSFET used in this work are shown in Fig. 1(a-b). The transistor measured in this work has a gate length ($L_G$)/Width (W) of 40/450 nm. For in-wafer device-to-device (D2D) variability analysis, the measurements are performed on 25 different dies across the wafer at room temperature (RT) and on 12 dies at higher temperatures (85 °C, and 125 °C). Agilent B1500 Semiconductor parameter analyzer is used for dc measurement. The variability in drain current ($I_D$) is analyzed for the following:
(1) Process variability (P): Device-to-device variability (D2D)

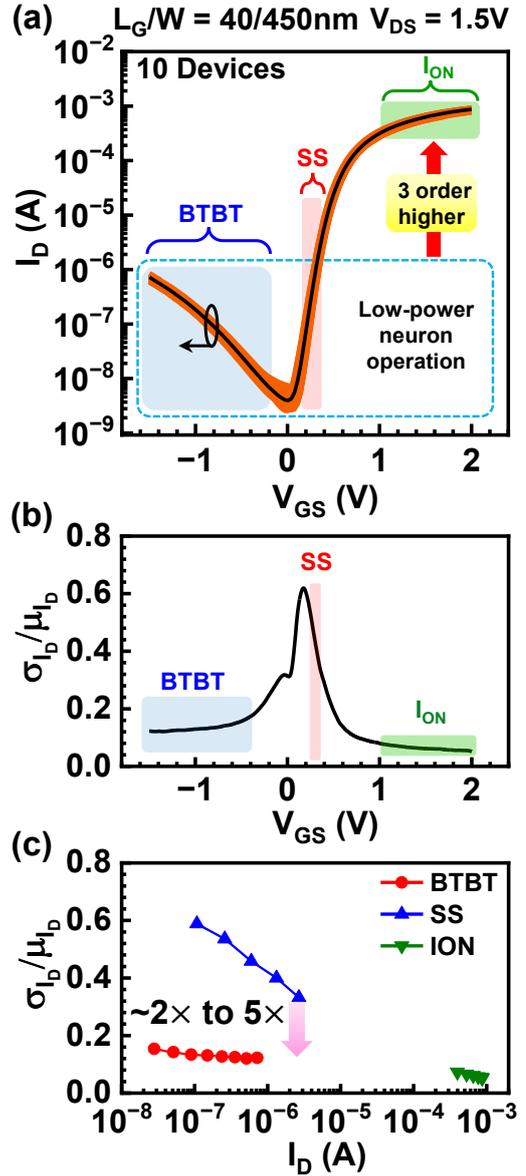

Figure 2 **Process-induced variability.** (a) Experimentally measured SOI MOSFET transfer characteristics for the nominal device ($L_G$/W = 40/450 nm) at $V_{DS}$ = 1.5 V for 25 devices measured across the wafer. (b) Calculated coefficient of variation ($\sigma_{I_D}/\mu_{I_D}$) as a function of the operating voltage. (c) Calculated $\sigma_{I_D}/\mu_{I_D}$ as a function of the drain current ($I_D$). It is observed that the $I_{ON}$ regime offers the lowest variability but dissipates higher power. Hence, the $I_{ON}$ regime is not suitable for low-power sources. The BTBT and SS regimes can have the same low power. However, BTBT has ~2× to 5× lower variability.

(2) Voltage variability (V): ΔV = 10% of operating voltage, and
(3) Temperature variability (T): Room temperature (RT), 85 °C, and 125 °C.

The threshold voltage ($V_T$) is extracted using the constant current method at $I_D = 5 \times 10^{-7}$ A.

### IV. RESULTS AND DISCUSSION

*A. Process Variability*

Fig. 2(a) shows the experimental transfer characteristic of the SOI MOSFET for 25 devices of the same dimension

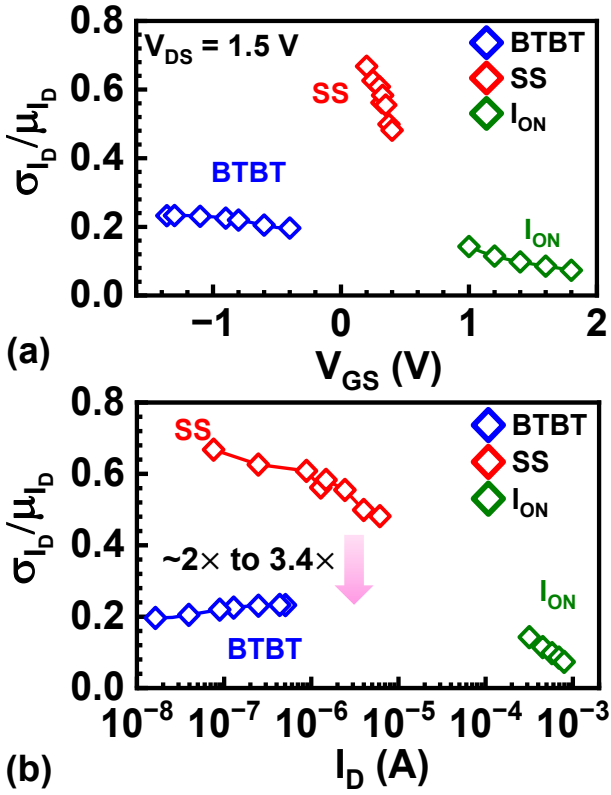

Figure 3 **Voltage variability.** Calculated coefficient of variation ($\sigma_{I_D}/\mu_{I_D}$) versus (a) the operating voltage and (b) the drain current ($I_D$) due to 10% voltage supply fluctuation in all three operating regimes. BTBT regimes offer ~2× to 3.4× lower variability than SS with voltage variation for the same $I_D$ operating range.

($L_G/W$ = 40/450 nm) measured across the wafer. The coefficient of variation ($\sigma_{I_D}/\mu_{I_D}$) is calculated by fitting a Gaussian distribution. The calculated $\sigma_{I_D}/\mu_{I_D}$ as a function of the operating voltage and the drain current is shown in Fig. 2(b) and Fig. 2(c), respectively. It is observed that the $I_{ON}$ regime offers the lowest variability but dissipates higher power. Hence, the $I_{ON}$ regime is not suitable for low-power sources. The BTBT and SS regimes can have similar power. However, BTBT offers ~2× to 5× lower variability (Fig. 2(c)) for the same drive currents. This is qualitatively consistent with the first-order analysis expectation (Table 1).

*B. Voltage Variability*

Besides D2D variability, the on-chip power supply voltage fluctuation is one of the major concerns in the circuit design. A 10% operating voltage ($V_{op}$) variation is considered to study the drain current variability under the voltage supply fluctuation ($\Delta V$ = 10% of $<V_{op}>$). Fig. 3 shows the calculated $\sigma_{I_D}/\mu_{I_D}$ as a function of the operating voltage (Fig. 3(a)) and the drive current (Fig. 3(b)) in the three operating regimes. It can be observed that the BTBT regime offer ~2× to 3.4× lower variability than the SS regime for the similar low-power operating range, while the $I_{ON}$ regime provides the least variability; however, it consumes high power. This is qualitatively consistent with the expectation from the first-order analysis presented earlier (Table I).

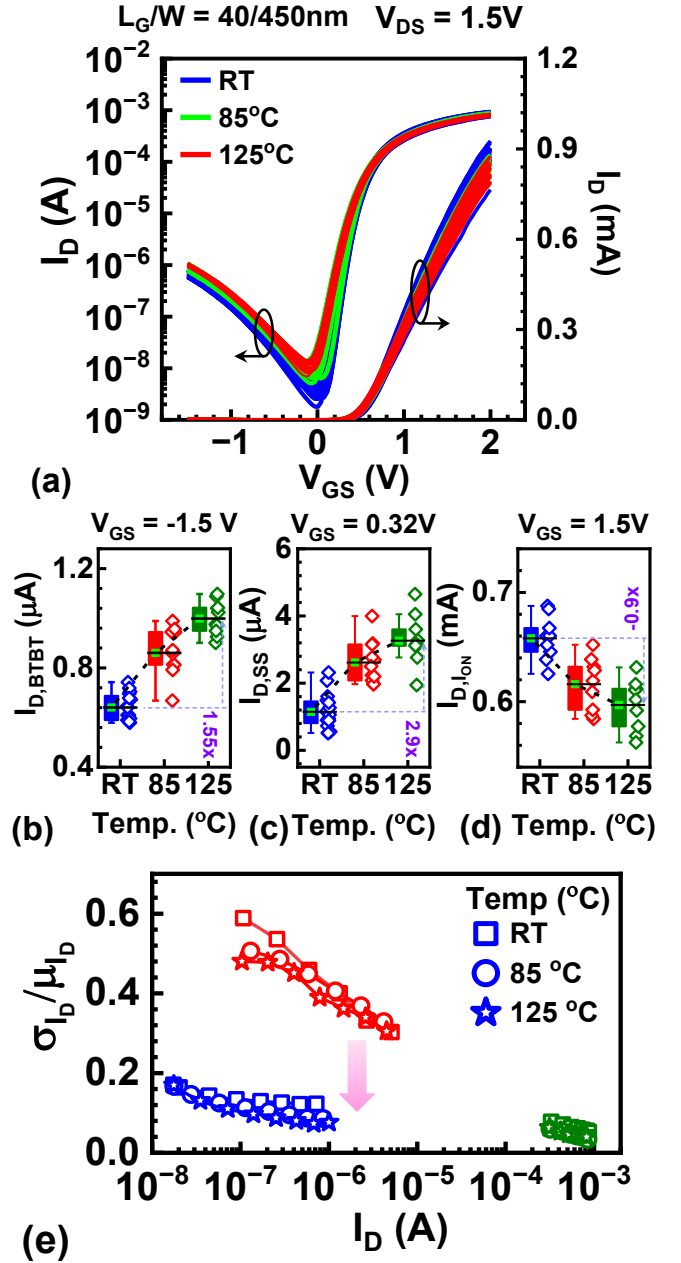

Figure 4 **Temperature variability.** (a) Measured transfer characteristics of the SOI transistor at RT, 85 °C, and 125 °C. The transistor dimension is $L_G/W$ = 40/450 nm. Box plot showing the extracted drain current in (b) BTBT, (c) SS, and (d) $I_{ON}$ regime at $V_{GS}$ of -1.5, 0.32, and 1.5 V, respectively, at fixed $V_{DS}$ of 1.5 V. The BTBT and SS regime shows an increase in current, while the current in the $I_{ON}$ regime decreases at higher temperatures due to an increase in phonon scattering. The $I_{D,BTBT}$ shows a ~1.87× lower increment than $I_{D,SS}$ at 125 °C compared to RT, implying that the BTBT regime has a low-temperature coefficient than the SS regime. (e) Calculated $\sigma_{I_D}/\mu_{I_D}$ as a function of $I_D$ comparison in the three operating regimes at different temperatures. It can be observed that the BTBT regime offers lower variability than the SS regime at all temperatures.

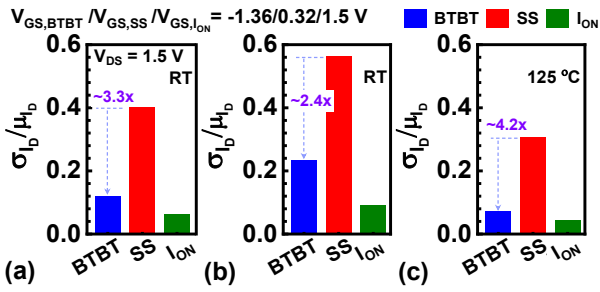

Figure 5 **PVT summary**. Comparison of coefficient of variation($\sigma_{I_D}/\mu_{I_D}$) for (a) process, (b) voltage, and (c) temperature variation in the BTBT, SS, and $I_{ON}$ operating regime at $V_{GS,BTBT}/V_{GS,SS}/V_{GS,I_{ON}}$ = -1.36/0.32/1.5 V at fixed $V_{DS}$ = 1.5 V. BTBT operating regime have shown lower $\sigma_{I_D}/\mu_{I_D}$ than the SS regime for the PVT variability.

TABLE II. COMPARISON OF POWER AND SIGMA/MEAN FOR THE PROCESS, VOLTAGE, AND TEMPERATURE VARIABILITY IN BTBT, SS, AND $I_{ON}$ OPERATING REGIME

| Comparison | | BTBT | SS | $I_{ON}$ | |
|---|---|---|---|---|---|
| Operating Power | | 0.1 μW @ $I_D$ = 112 nA | 0.1 μW @ $I_D$ = 359 nA | 870 μW @ $I_D$ = 580 μA | |
| Variability parameters | | $\left(\frac{\sigma}{\mu}\right)_{I_{D,BTBT}}$ | $\left(\frac{\sigma}{\mu}\right)_{I_{D,SS}}$ | $\left(\frac{\sigma}{\mu}\right)_{I_{D,ION}}$ | Ratio: $\frac{(\sigma/\mu)_{SS}}{(\sigma/\mu)_{BTBT}}$ |
| Process variation | $L_G$ = 40 nm | 0.12 | 0.4 | 0.061 | 3.3 |
| Voltage variation | $\Delta V$ = 10% $<V>$ | 0.23 | 0.56 | 0.09 | 2.4 |
| Temperature variation | 125 °C | 0.072 | 0.3 | 0.042 | 4.2 |
| Total (σ/μ: P V T) | | 0.272 | 0.75 | 0.12 | 2.77 |

*C. Temperature Variability*

The manufactured transistors are intended to support a wide range of applications. Depending on the applications to circumvent circuit malfunction, the transistors must function reliably within a specified temperature range. Hence, we investigated the impact of elevated temperatures (85 °C and 125 °C) on the transistor performance in all three operating regimes.

The measured transfer characteristics for the transistor of $L_G/W$ = 40/450 nm at room temperature (RT), 85 °C, and 125 °C are shown in Fig. 4(a). It is evident that the temperature affects the transistor's performance. To analyze the variation in drive current ($I_D$), the $I_D$ in BTBT, SS, and $I_{ON}$ operating regimes at $V_{GS}$ of -1.5, 0.32, and 1.5 V is extracted, as shown in Fig. 4(b)-(d). The current in BTBT and SS regimes shows an increment by ~1.55× and ~2.9×, respectively, whereas the $I_{ON}$ regime shows a decrement by -0.9× at 125 °C compared to RT. As shown, the $I_{D,BTBT}$ shows a ~1.87× lower increment than $I_{D,SS}$ at 125 °C as compared to RT, implying that the BTBT regime has a lower temperature coefficient than the SS regime (Fig. 4(b)-(c)).

The calculated $\sigma_{I_D}/\mu_{I_D}$ as a function of the $I_D$ comparison in the three operating regimes with temperature dependence is shown in Fig. 4(e). It can be observed that $I_{D,BTBT}$ offers lower $\sigma_{I_D}/\mu_{I_D}$ than $I_{D,SS}$ at all the temperature ranges.

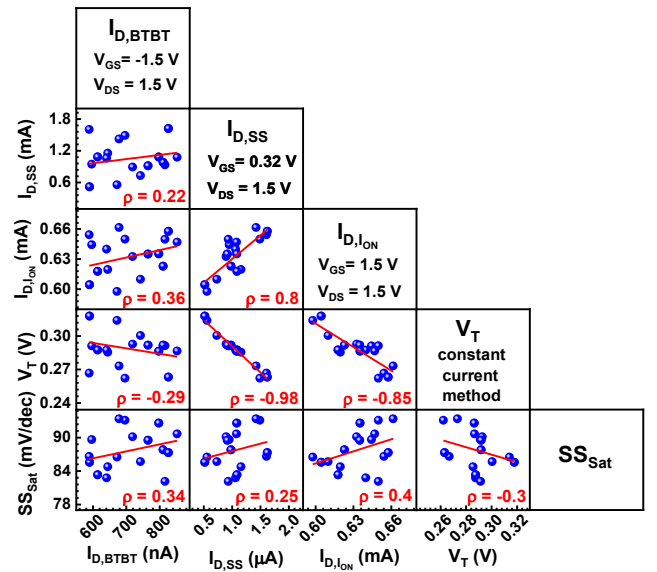

Figure 6 **Correlation analysis**. The Q-Q plot for the measured parameters $I_{D,BTBT}$, $I_{D,SS}$, $I_{D,I_{ON}}$, $V_T$, and $SS_{sat}$ and their correlations. It is observed that the BTBT variation is uncorrelated with mutually correlated SS & $I_{ON}$ operation – indicating its different origin from the mechanism and location perspectives.

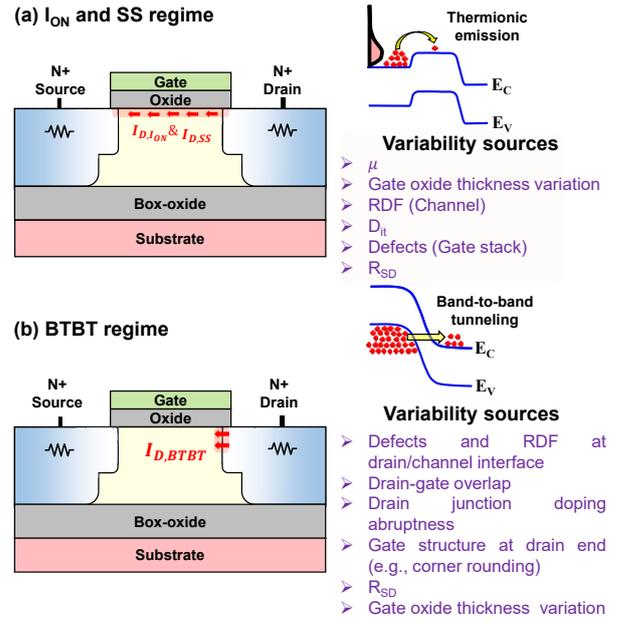

Figure 7 **Variability source.** SOI MOSFET schematic showing the source of variability in (a) the SS and $I_{ON}$ regime and (b) the BTBT regime. The SS and $I_{ON}$ regime current depend upon the transport in the channel, which is affected by the channel mobility degradation ($\mu$), gate oxide thickness variation (OTV), random dopant fluctuation (RDF) in channel region, interface traps ($D_{it}$), defects in gate stack, and source/drain resistance ($R_{SD}$). The BTBT current variability is mainly due to the defects and RDF at the drain/channel interface, variation in the (i) drain-gate overlap, (ii) drain junction doping abruptness, (iii) gate structure at the drain end (e.g., corner rounding, etc.), (iv) $R_{SD}$ and (vi) OTV, causing a change in the tunneling distance/barrier at the junction.

*D. Variability summary*

To summarize, a comparison is shown for the process, voltage, and temperature variation in the drain current ($I_D$) in the three operating regions at a gate-to-source voltage ($V_{GS}$): (1) $V_{GS,BTBT}$ = -1.36 V, (2) $V_{GS,SS}$ = 0.32 V and (3) $V_{GS,I_{ON}}$ = 1.5 V at fixed $V_{DS}$ of 1.5 V (Fig. 5). The calculated values are tabulated in Table II for power comparison and the effects of PVT variability on $I_D$. Due to its higher power consumption, the $I_{ON}$ regime is not preferred over the BTBT and SS regimes. Among the BTBT and SS regimes, BTBT offers a significantly lower $\sigma_{I_D}/\mu_{I_D}$ over the SS operating regime for the PVT variability. Cumulatively, BTBT shows a ~3× variability reduction than the SS regime for PVT variability (Table II).

*E. Correlation analysis*

To have a better comprehensive correlation picture between different device parameters $I_{D,BTBT}, I_{D,SS}, I_{D,I_{ON}}, V_T$, and $SS_{sat}$, we have performed the correlation analysis by Q-Q plot (Fig. 6). The $I_{D,BTBT}$-$V_T$ shows a minimal correlation (ρ = -0.29), whereas $I_{D,SS}$-$V_T$ shows a very high correlation (ρ = -0.98). $I_{D,I_{ON}}$-$V_T$ shows a lower correlation than $I_{D,SS}$-$V_T$. It implies that the SS operation regime suffers strongly from device variability due to its exponential dependence on $V_T$. We observe that the BTBT variation is uncorrelated with the mutually correlated SS & $I_{ON}$ operation – indicating its different origin from the mechanism and location perspectives.

*F. Discussion*

The $I_{ON}$ and SS currents depend upon current transport in the channel (Fig. 7a). In contrast, the BTBT currents depend upon current transport in the drain-channel junction right under the gate (Fig. 7b). These fairly different locations are reflected strongly in the mutual correlation between SS and $I_{ON}$ and their absence of correlation with the BTBT from a process perspective – even though all these effects are measured for the same set of devices. The variability in the BTBT current is mainly due to variation in the drain-gate overlap, gate oxide thickness variation (OTV), defects and random dopant fluctuation at the drain/channel interface, doping abruptness at drain junction, gate structure at the drain end (e.g., corner rounding, etc.), and $R_{SD}$, causing a change in the tunneling distance/barrier at the junction (Fig. 7). The BTBT current has lower variability as these variability sources are less dominant. Although BTBT shows low variability compared to Ion and SS regimes, it is difficult to define the percent of each variability source quantitatively. For voltage and temperature, the difference in the mechanism produces a difference in sensitivity to these parameters. The low variability in $I_{ON}$ comes at the cost of a higher current. As the primary interest is in low current mechanisms, for the same low current, BTBT shows weaker temperature and voltage dependence as expected qualitatively from the first-order analysis.

## V. CONCLUSION

We have demonstrated that the BTBT regime is promising for low-power neurons. In addition to the area and energy efficiency (fJ/Spike) demonstrated earlier, here we have experimentally demonstrated that BTBT offers lower PVT variation (~ 3 ×) compared to the SS regime at the same current levels. The difference in location of the BTBT current shows in the absence of correlation with SS/$I_{ON}$, which are mutually well-correlated. Our study is an important step in the development of neuromorphic hardware design using BTBT-based neurons.